\documentclass[a4paper,12pt]{article}
\usepackage{epsfig}
\usepackage{graphicx}
\begin{document}
\begin{titlepage}


\vfill

{\bf\LARGE
\begin{center}
           Sensitivity of the EDELWEISS WIMP search
           to spin-dependent interactions
\end{center}}

\vfill

\begin{center}
{\large The EDELWEISS Collaboration:} \\
A.~Benoit$^{1}$,
L.~Berg\'e$^{2}$,
J.~Bl\"umer$^{3,4}$,
A.~Broniatowski$^{2}$,
B.~Censier$^{2}$,
L.~Chabert$^{5}$,
M.~Chapellier$^{6}$,
G.~Chardin$^{7}$,
S.~Collin$^{2}$,
M.~De~J\'esus$^{5}$,
H.~Deschamps$^{7}$,
P.~Di~Stefano$^{5}$,
Y.~Dolgorouky$^{2}$,
D.~Drain$^{5}$,
L.~Dumoulin$^{2}$,
K.~Eitel$^{4}$,
M.~Fesquet$^{7}$,
S.~Fiorucci$^{7}$,
J.~Gascon$^{5}$,
G.~Gerbier$^{7}$,
C.~Goldbach$^{8}$,
M.~Gros$^{7}$,
R.~Gumbsheimer$^{4}$,
M.~Horn$^{4}$,
A.~Juillard$^{2}$,
A.~de~Lesquen$^{7}$,
M.~Luca$^{5}$,
J.~Mallet$^{7}$,
S.~Marnieros$^{2}$,
L.~Mosca$^{7}$,
X.-F.~Navick$^{7}$,
G.~Nollez$^{8}$,
P.~Pari$^{6}$,
M.~Razeti$^{5}$,
V.~Sanglard$^{5}$,
L.~Schoeffel$^{7}$,
M.~Stern$^{5}$,
V.~Villar$^{7}$
\end{center}

{\scriptsize\noindent
$^{1}$Centre de Recherche sur les Tr\`es Basses Temp\'eratures,
    SPM-CNRS, BP 166, 38042 Grenoble, France\\
$^{2}$Centre de Spectroscopie Nucl\'eaire et de Spectroscopie de Masse,
     IN2P3-CNRS, Universit\'e Paris XI,
    b\^at~108, 91405 Orsay, France\\
$^{3}$Institut f\"ur Experimentelle Kernphysik, Universit\"at
      Karlsruhe (TH), Gaedestr.~1,\\
      76128~Karlsruhe, Germany\\
$^{4}$Forschungszentrum Karlsruhe, Institut f\"ur Kernphysik,
     Postfach 3640, 76021 Karlsruhe, Germany\\
$^{5}$Institut de Physique Nucl\'eaire de Lyon-UCBL, IN2P3-CNRS,
     4 rue Enrico Fermi,\\
     69622 Villeurbanne~Cedex, France\\
$^{6}$CEA, Centre d'\'Etudes Nucl\'eaires de Saclay,
     DSM/DRECAM, 91191 Gif-sur-Yvette Cedex, France\\
$^{7}$CEA, Centre d'\'Etudes Nucl\'eaires de Saclay,
     DSM/DAPNIA, 91191 Gif-sur-Yvette Cedex, France\\
$^{8}$Institut d'Astrophysique de Paris, INSU-CNRS,
     98 bis Bd Arago, 75014 Paris, France
}

\vfill

\begin{center}{\large\bf Abstract}\end{center}

The EDELWEISS collaboration is searching for WIMP dark matter using natural
Ge cryogenic detectors. The whole data set of the first phase of the
 experiment contains a fiducial exposure of 4.8 kg.day 
on $^{73}$Ge, the naturally present (7.8\%),
 high-spin Ge isotope.
The sensitivity of the experiment to the spin-dependent WIMP-nucleon 
interactions is evaluated using the model-independent framework
proposed by Tovey {\it et al.}~\cite{tovey}. It is shown that
the EDELWEISS sensitivity for the WIMP-neutron coupling is competitive
when compared with results of other spin-sensitive WIMP dark matter
experiments. The current experimental limits lie however two orders of 
magnitude higher than the most optimistic SUSY models.
\vfill

\begin{center}
{\bf PACS classification codes:}
95.35.+d, 
14.80.Ly, 
98.80.Es, 
29.40.Wk. 
\end{center}

\end{titlepage}


\noindent{\large\bf Introduction} \\

In direct searches for
the Galactic cold dark matter in the form of Weakly Interacting Massive
Particles (WIMPs),
the experimental signature is the observation of nuclear recoils
induced by WIMP scattering off nuclei in a terrestrial
detector. If the WIMP is assumed to be the neutralino, the lightest
supersymmetric particle,
two types of couplings with matter have to be
considered: scalar (spin-independent) and axial-vector 
(spin-dependent)~\cite{jung}. 

For sufficiently heavy nuclei, the spin-independent interaction is expected
to give the dominant contribution~\cite{jung}, due to the coherent
enhancement approximately proportional to $A^2$, the square of the mass number.
Experimental results are thus most often given as exclusion curves for the
spin-independent WIMP-nucleon diffusion cross-section (see for 
instance~\cite{{igex},{edw1},{CDMS},{cresst}} 
and references therein).  

For spin-dependent couplings, this enhancement is not present due to
cancellation effects between paired nucleons within the
nucleus. For example, in the single-particle shell model~\cite{good} 
the nuclear 
spin comes from
the unpaired proton or neutron.
Even with high-spin nuclei the sensitivity
of direct detection experiments on the WIMP-nucleon cross-section is orders of
magnitude lower than for the spin-independent case 
(see Refs.~\cite{giuliani} and~\cite{savage} for
recent reviews). For axial-vector coupling, indirect detection
experiments searching for
high-energy neutrinos produced by WIMP annihilations in the Sun provide
much higher sensitivity than direct detection ones~\cite{savage}: 
the capture rate of WIMPs
in the Sun core is governed by the scattering rate on
protons (low mass, high specific spin).

For spin-dependent interactions, comparisons between 
WIMP-nucleon cross-section limits given by
various experiments is problematic: the comparison of results
obtained using targets with either an unpaired neutron or an unpaired
proton relies on the assumptions concerning the ratio of the WIMP-neutron
to the WIMP-proton cross-sections. This is particularly obvious
in the framework of the simple single-particle model.
Although it is expected that this ratio should be ${\cal O}(1)$ in minimal
supersymmetric models~\cite{bednyakov2}, Tovey 
{\it et al.}~\cite{tovey} have proposed a method to extract limits on 
the spin-dependent interaction in the more general case where there
is no constraint on this ratio.
For a given WIMP mass, exclusion curves are obtained
in the plane of the
effective WIMP-neutron and WIMP-proton couplings.
The same type of exclusion
curves has been computed very recently~\cite{savage} using a more 
rigorous formalism. Evaluations of current results of spin-dependent
WIMP searches have
been given using both techniques~\cite{{giuliani},{savage}}.

The EDELWEISS collaboration searches for WIMP dark matter using natural
Ge cryogenic detectors~\cite{{edw0},{edw1},{edw2}}. 
The whole data set of the first phase of the
experiment contains a fiducial exposure of 4.8 kg.day 
on $^{73}$Ge, the naturally present (7.8\%),
high-spin Ge isotope.
The aim of this paper is to compare, for
the spin-dependent interaction and using the model-independent framework
of Ref.~\cite{tovey}, EDELWEISS limits with those of the most representative
high spin target experiments. It will be shown that, despite the
very low $^{73}$Ge content of natural Ge, the EDELWEISS sensitivity for the
WIMP-neutron coupling is competitive.

\mbox{}\\\noindent{\large\bf Experimental data} \\

The EDELWEISS experiment is set in the Laboratoire Souterrain 
de Modane (LSM) in
the Fr\'ejus tunnel connecting France and Italy. Detectors used by the
experiment are 320~g Ge phonon and ionization cryogenic detectors 
placed in a dilution 
cryostat at a regulated temperature of 17~mK. The experimental setup is
described elsewhere~\cite{{edw1},{edw0}}. 

The main characteristic of the EDELWEISS detectors is the simultaneous measurement
of the phonon and the ionization signals. 
The ionization signal is measured by  
Al electrodes sputtered on each side of the crystal and the phonon signal
by a NTD heat sensor glued onto one electrode. The measurement of both signals 
provides a very good event-by-event discrimination between nuclear and electronic 
recoils, typically more than 99.9$\%$ above 15~keV. More details on the detectors 
and their performances can be found in Ref.~\cite{edw2}. 
Following the first published data~\cite{{edw1},{edw0}}, 
a new set of three detectors was operated 
in the cryostat, all of them having either a Ge or Si amorphous 
layer for better 
charge collection.

Between 2000 and 2003, four physics runs have been 
performed with different 
detectors, trigger conditions and efficiencies. All these runs have been
re-analysed with an uniform analysis threshold of 15 keV and using the 
efficiency versus recoil energy function of each run.
A detailed presentation of the experimental data and a thorough 
discussion of their analysis is given in Ref.~\cite{edw4}. Only the most 
relevant features are summarized hereafter.

In 2000 and 2002, 13.6~kg.day (after the fiducial volume cut) were 
accumulated with two detectors~\cite{{edw1},{edw0}}. Two events 
compatible with nuclear recoils have been recorded in 2000 above the 
new analysis threshold of 15 keV, and two in 2002.
In 2003 two other runs
have been performed with new detectors but with two different triggers. 
In the first run, the trigger was the fast ionization signal (run 2003i, 
fiducial exposure 25.7~kg.day). 
In the second run, the trigger was the phonon signal (run 2003p,
fiducial exposure 22.7~kg.day).
This latter trigger condition 
improves the efficiency at low energy: the baseline resolution is better
on the phonon channel
and the full recoil energy deposition
$E_R$ is
recorded, instead of the quenched fraction
$E_I = Q.E_R$ in the ionization channel.
In the runs 2003i and 2003p, the numbers of observed events
compatible with nuclear recoils above the analysis 
threshold are 17 and 19, respectively.
This apparent increase of the raw nuclear recoil rates in the
new 2003 runs is explained by the significant increase in efficiency 
at low energy in the new data set (see Ref.~\cite{edw4} for details). 
Only three events lie in the energy
interval from 30 to 100 keV, a result consistent with the previous
data sets~\cite{{edw1},{edw0}}.
The entire set of data for the EDELWEISS-I experiment 
consists therefore of a fiducial exposure of 62.0~kg.day with a 
total of 40 events compatible 
with nuclear recoils 
above 15~keV. 
Conservatively considering all these events as WIMP interactions, 
a 90~$\%$~C.L. upper limit on the WIMP-nucleon cross-section as a 
function of the WIMP 
mass is derived with the Optimum Interval Method~\cite{yellin}. 
 
For the spin-dependent coupling, one must consider the exposure on $^{73}$Ge (4.8~kg.day) 
and a specific form factor. Here, the form factor of 
Dimitrov {\it et al.}~\cite{dimitrov} 
is used, with the usual approximation that the isoscalar, 
isovector and interference form factors
are identical
in order to make the form factor independent of the WIMP-nucleon
couplings~\cite{tovey}. The resulting uncertainty on the cross-section is
within $\pm$15\% up to a WIMP mass of 1 TeV/c$^2$. 
Other calculations of spin-dependent 
scattering on $^{73}$Ge are
briefly presented hereafter. The resulting 90~$\%$~C.L. limits 
on the WIMP-nucleus cross-section $\sigma_A$ 
for WIMP masses between 20~GeV/c$^2$ and 1~TeV/c$^2$ are listed 
in table~\ref{resultat}. These limits are conservative as they neglect any
contribution from the spin-independent coupling.  
 
\mbox{}\\\noindent{\large\bf Model-independent Exclusion Limits}\\

For spin-dependent interactions the WIMP-nucleus cross-section $\sigma_A$
at zero momentum transfer can be approximated by the expression
(see for instance~\cite{jung}):
$$ \sigma_A={32\over \pi}G_F^2\mu_A^2(a_p\left<S_p\right>+a_n\left<S_n\right>)^2{J+1\over J},
\eqno (1)$$
where $G_F$ is the Fermi coupling constant,
 $\mu_A$ the WIMP-nucleus reduced mass,
$a_{p,n}$ the effective WIMP-proton(neutron) couplings, $\left<S_{p,n}\right>$ the 
expectation values of the proton(neutron) spins within the nucleus and $J$ the
total nuclear spin. WIMP-nucleon cross-sections $\sigma^{lim(A)}_{p,n}$ in
the limit $a_{n,p}=0$ respectively, are defined as~\cite{tovey}:
$$\sigma^{lim(A)}_{p,n}={3\over 4}{J\over J+1}{\mu_{p,n}^2\over \mu_A^2}
{\sigma_A\over \left<S_{p,n}\right>^2},\eqno (2)$$
where $\mu_{p,n}$ is the WIMP-proton(neutron) reduced mass and $\sigma_A$ the
WIMP-nucleus cross-section limit (at 90\% CL) deduced from the experiment. 

It is shown in~\cite{tovey} that the {\it allowed} values of $a_p$ and $a_n$
for a particular WIMP mass 
obey the inequality:
$$\left({a_p\over \sqrt{\sigma^{lim(A)}_p}}\pm{a_n\over \sqrt{\sigma^{lim(A)}_n}}\right)^2
\leq {\pi\over 24G_F^2\mu_p^2}.\eqno (3)$$
 
The sign between parentheses is that of ${\left<S_p\right>\over \left<S_n\right>}$. Equation (3)
defines two parallel straight lines in the $(a_n,a_p)$ plane, the slope of
which is ${-\left<S_n\right>\over \left<S_p\right>}$.
The allowed values of $a_p$ and $a_n$ are within
the band defined by these two lines (while in~\cite{savage} an extremely
elongated ellipse is found). For experiments with two active nuclei
with different ${\left<S_p\right>\over \left<S_n\right>}$ ratios 
the combination of the two bands gives rise to a closed elliptical contour.

Nuclear spin structure calculations have been recently reviewed by 
Bednyakov and \v Simkovic~\cite{bednyakov}: for $^{73}$Ge the two most
comprehensive spin structure analyses are from Ressel et al.~\cite{ressel}
and from Dimitrov et al.~\cite{dimitrov} (see table 2). The corresponding 
form factors are very similar. In these calculations the neutron spin
${\left<S_n\right>}$ varies by $\sim$20\% depending on the nuclear stucture 
model. The value of the ratio 
${\left<S_n\right>\over \left<S_p\right>}$ is model dependent, 
but is always much greater than unity. Thus the odd-N, even-Z 
nucleus $^{73}$Ge nucleus is mainly 
sensitive to $a_n$ only, and in that sense is complementary to other 
detectors made out of odd-Z material such as $^{23}$Na, $^{127}$I, $^7$Li,
$^{19}$F or $^{27}$Al.

Table 1 gives the values of the spin-dependent WIMP-nucleon cross-sections
deduced from our experimental $\sigma_A$ values
using eq. (2) and the $\left<S_{p,n}\right>$ taken
from~\cite{dimitrov}.

Plots of the WIMP-nucleon cross-sections $\sigma^{lim(A)}_{p,n}$ 
versus the
WIMP mass are shown in Fig. 1 and 2.
Figures 3 and 4 show the regions in the 
$(a_n,a_p)$ plane allowed by
EDELWEISS, using eq. (3) and the experimental values of table~1,
for two illustrative WIMP masses $M_{\chi}=$ 50 and 500 GeV/c$^2$.
A comparison is made
with other experiments representative
of various techniques: NaI scintillators (NAIAD experiment~\cite{ahmed}),
fluored bolometers~(Kamioka LiF~\cite{miuchi}), sapphire 
bolometers (CRESST-I~\cite{anglo}), freon droplets 
(SIMPLE~\cite{girard}, PICASSO~\cite{picasso})
and the other natural Ge cryogenic
experiment (CDMS-II~\cite{CDMS}).
A more exhaustive comparison can be found in the review of
Giuliani~\cite{giuliani}. All experiments use the same dark 
matter halo model 
to extract $\sigma_A$. Form factors, values
of  $\left<S_{p,n}\right>$ specific to each nucleus 
and WIMP-nucleon cross-sections $\sigma^{lim(A)}_{p,n}$ are given 
in~\cite{ahmed},~\cite{miuchi},~\cite{girard} and~\cite{picasso} . 

For the CRESST-I (sapphire) experiment,
relevant cross-sections can be deduced from
the published data using $\left<S_{p,n}\right>$ 
values from~\cite{engel} for $^{27}$Al
(neglecting the $^{16}$O spin sensitivity). For the CDMS experiment
limits on spin-dependent nucleon cross-sections are derived in 
Ref.~\cite{savage}, including the coupling dependency in the form factor
rather than taking the isoscalar, isovector and interference terms 
to be identical.
For consistency, $\sigma^{lim(A)}_{p,n}$ are recomputed 
from the CDMS data
(no event in 52.6 kg.day raw exposure in Ge detectors over 10 to 100 keV
recoil energy) using the experimental efficiency quoted in Ref.~\cite{savage},
form factor and $\left<S_{p,n}\right>$ values from Ref.~\cite{dimitrov}.
The differences in the $a_n$ allowed regions defined by the two calculations 
are insignificant in the limit $a_p=0$.

The DAMA/Xe experiment~\cite{bernabei} makes use of a $^{129}$Xe target, 
an other example of odd-N, even-Z nucleus, as $^{73}$Ge. 
The $\sigma^{lim(A)}_p$ can be calculated from the reported 
$\sigma^{lim(A)}_n$ cross-section.
However this result deserves a special comment.
The value
of the quenching factor for liquid Xenon is taken 
in~\cite{bernabei} as $Q=$ 0.44,
i.e. more than a factor
of 2 greater than other measured
values~\cite{{arneodo},{akimov}}. 
Adopting these more recent and precise values shifts
the nuclear recoil energy threshold of the DAMA/Xe experiment
from 30 keV to at least 60 keV. For
$M_{\chi}=$ 50 GeV/c$^2$ the WIMP event rate at the threshold is 
then divided by a factor of 15 and even more if the fast decrease 
of the form 
factor with recoil energy is taken into account. The cross-section limit is
underestimated by the same factor. This considerable uncertainty on the actual 
recoil energy scale does not allow any reliable comparison and the published
results of DAMA/Xe are not shown on the figure.

Fig. 3 and 4 show that cryogenic natural Ge experiments 
such as EDELWEISS and CDMS give the most stringent limits on $a_n$,
at roughly the same level as the $a_p$ limits given by odd-Z detectors.
The low $^{73}$Ge content of natural Ge is balanced by the
very low level of nuclear recoil backgrounds achieved in the cryogenic
detectors and by the 
high neutron nuclear spin value of $^{73}$Ge.
The allowed region of minimal extension in the $(a_n,a_p)$ plane is 
given by the combination of experiments with respectively 
high $\left<S_n\right>$ 
and high $\left<S_p\right>$ nuclei. The best limit on
$a_n$ is presently set by the CDMS experiment~\cite{CDMS}.
As shown in Ref.~\cite{savage} the limit on $a_p$ set by the neutrino observatories 
Super-K~\cite{superK} and Baksan~\cite{baksan} is more than a factor of 10
better than current direct detection results.

\mbox{}\\\noindent{\large\bf Conclusions}\\

For a WIMP mass between 50 and 500 Gev/c$^2$, 
current limits given by direct detection
experiments are 
$|a_{n,p}|<$ few units, or equivalently  $\sigma_{p,n}^{SD}<$ few pb.
The maximal values given by SUSY models lie  
 two orders of magnitude
lower (see for instance~\cite{{bednyakov2},{ellis},{profumo}}).
Only a new generation of experiments
designed to gain two orders
of magnitude in sensitivity will be able to reach the SUSY
predictions for $\sigma_{p,n}^{SD}$.
EDELWEISS-II is one of these forthcoming experiments~\cite{edw5}.

In the experimentally constrained versions of Supersymmetry, the neutralino 
is not a pure gaugino and one expects $|a_p|/|a_n|\sim{\cal O}(1)$
(for example see Ref.~\cite{bednyakov2}). To go beyond these constraints
and fix limits on $a_p$ by the direct detection technique,
high sensitivity experiments, using efficient background rejection
techniques and high $\left<S_p\right>$ nuclear targets are needed. 
A multi nuclear target experiment,
with powerful background discrimination capability and complementary 
proton vs neutron spin values would constitute an interesting new approach.
Setting decisive constraints on
the SUSY parameters
for spin-dependent WIMP-nucleon interactions  remains
a considerable experimental challenge for direct detection experiments.

\mbox{}\\\noindent{\large\bf Acknowledgments }\\

It is a pleasure to thank Franco Giuliani and Chris Savage for 
stimulating discussions. The 
help of the technical staff of the Laboratoire Souterrain de Modane 
and of the participating laboratories is gratefully acknowledged.
This work has been partially funded by the EEC Network program under
contract HPRN-CT-2002-00322.

\newpage

\begin{table}[h]
\begin{center}
\begin{tabular}[h]
{|c||c|c|c|}
\hline
$M_{\chi}$(GeV/c$^2$)&$\sigma_A$(pb)&$\sigma^{lim(A)}_p$(pb)&
$\sigma^{lim(A)}_n$(pb)\\
\hline
20&769.&1759.&11.1 \\
\hline
30&304.&395.&2.50 \\
\hline
40&201.&181.&1.14 \\
\hline
50&192.&133.& 0.845\\
\hline
60&199.&114.&0.721 \\
\hline
80&238.&103.&0.651 \\
\hline
100&291. &104.&0.661 \\
\hline
200&648. &149.&0.945\\
\hline
400&1482.&262.&1.66 \\
\hline
500&1917.&319.&2.02 \\
\hline
600&2358.&377.&2.39 \\
\hline
800&3247.&494.&3.13 \\
\hline
1000&4140.&611.&3.87 \\
\hline
\end{tabular}
\end{center}
\caption{
         \label{resultat}
         \textit{ 
Values of the WIMP-nucleus cross-section limit $\sigma_A$ (at 90\% CL)
deduced from the experiment. The corresponding $\sigma^{lim(A)}_{p,n}$
cross-sections (eq. (2)) are calculated using $\left<S_{p,n}\right>$ values 
from~\cite{dimitrov}.
                 }
         }    
\end{table}

\begin{table}[h]
\begin{center}
\begin{tabular}[h]
{|c||c|c|}
\hline
 &$\left<S_p\right>$&$\left<S_n\right>$\\
\hline
Ressel {\it et al.}~\cite{ressel}&0.011 &0.468\\
unquenched values& &\\
\hline
Ressel {\it et al.}~\cite{ressel}&0.009 &0.372\\
quenched values& &\\
\hline
Dimitrov {\it et al.}~\cite{dimitrov}&0.030 &0.378\\
\hline
\end{tabular}
\end{center}
\caption{
         \label{vfidt}
         \textit{ 
Spin values for $^{73}$Ge ($J=9/2$). See~\cite{bednyakov} for a critical
evaluation of the various calculations.
                 }
         }    
\end{table}

\newpage

\begin{figure}[tbh]
\epsfig{file=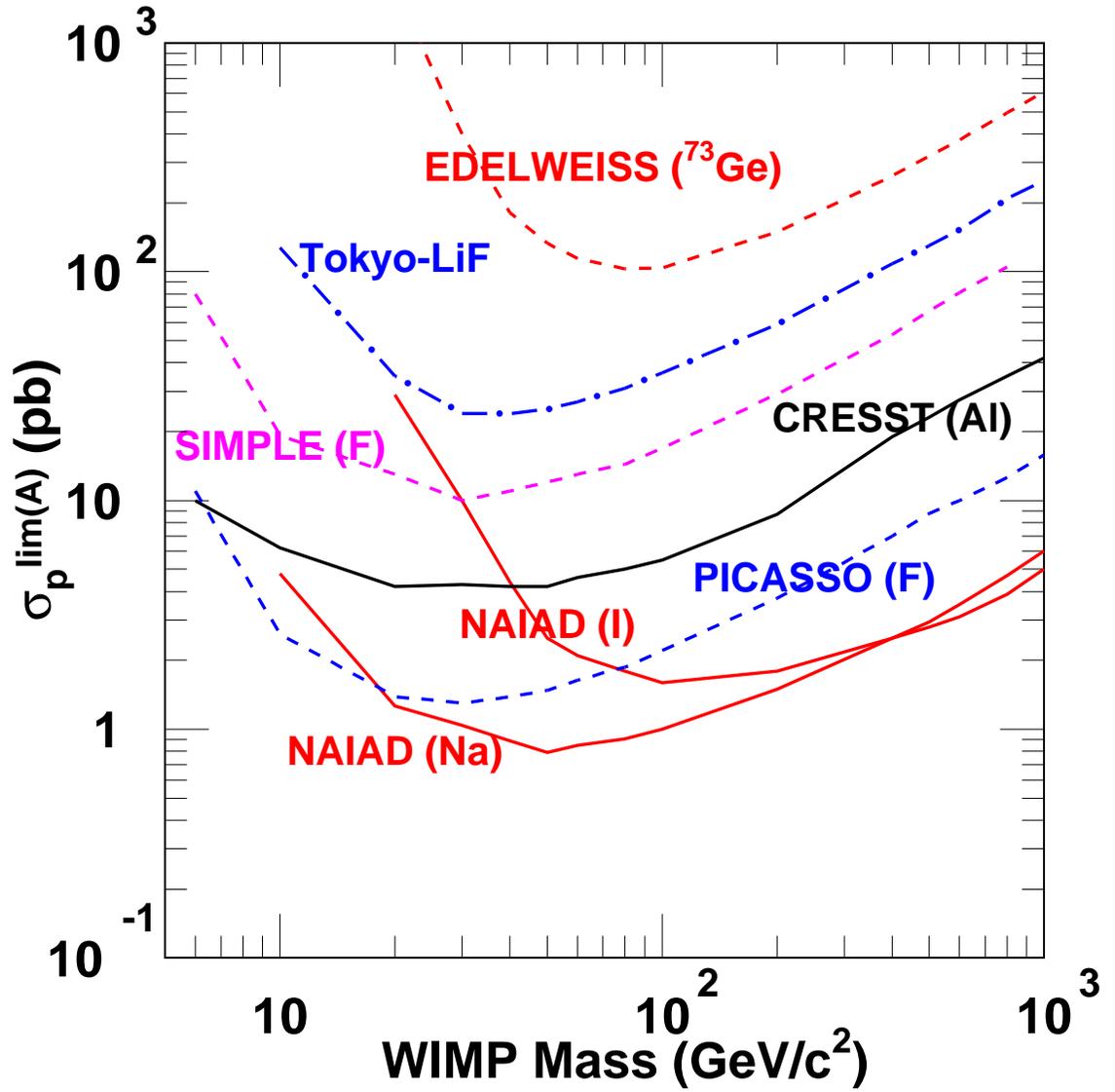
,height=17cm}
\caption[]{
$\sigma^{lim(A)}_{p}$ versus WIMP mass for the various experiments quoted
in the text (90\% CL limits).
          } 
\label{fig-1} 
\end{figure}

\newpage

\begin{figure}[tbh]
\epsfig{file=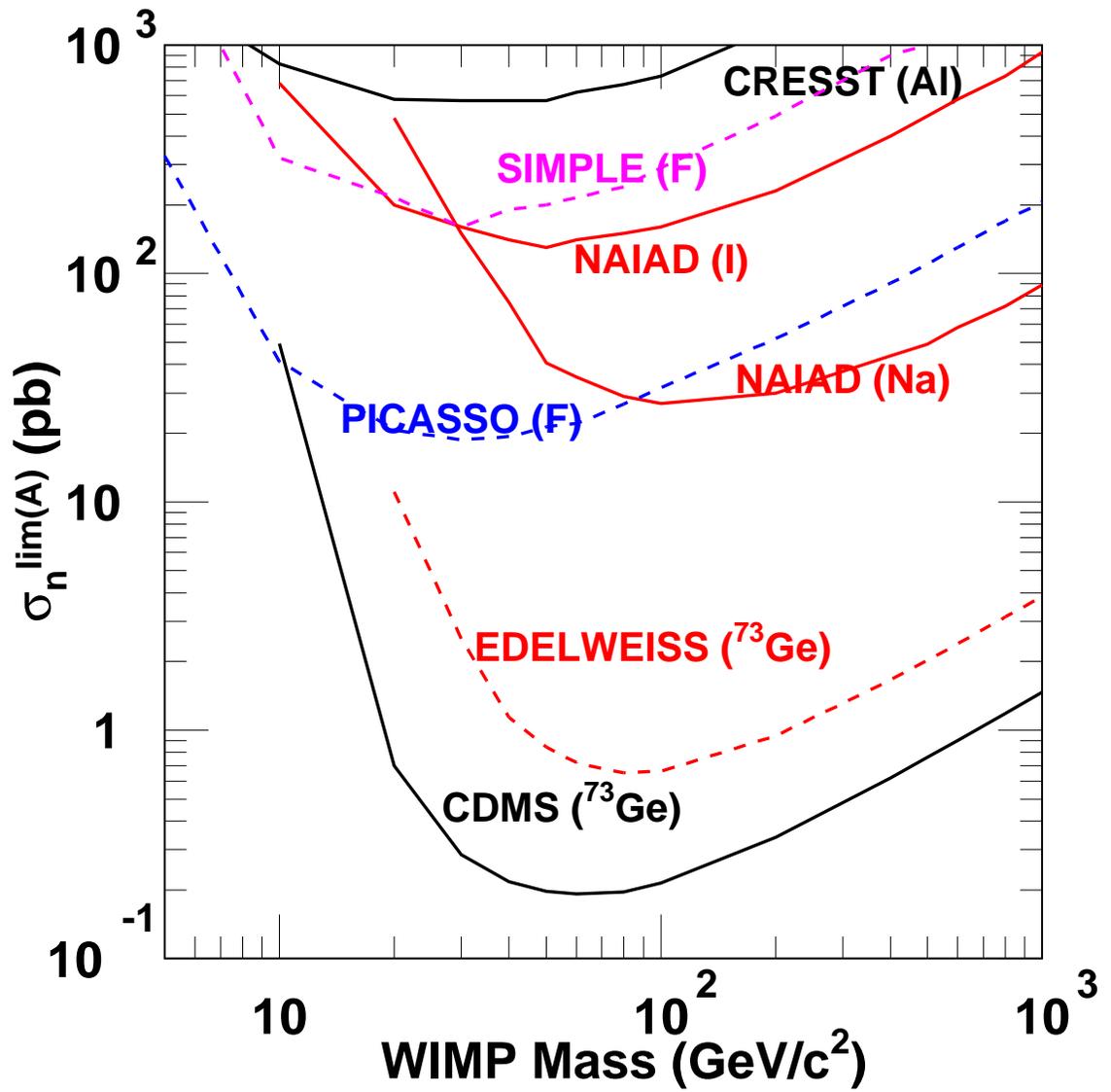
,height=17cm}
\caption[]{
The same as Fig. 1 for $\sigma^{lim(A)}_{n}$.
          } 
\label{fig-2} 
\end{figure}
 
\newpage

\begin{figure}[tbh]
\epsfig{file=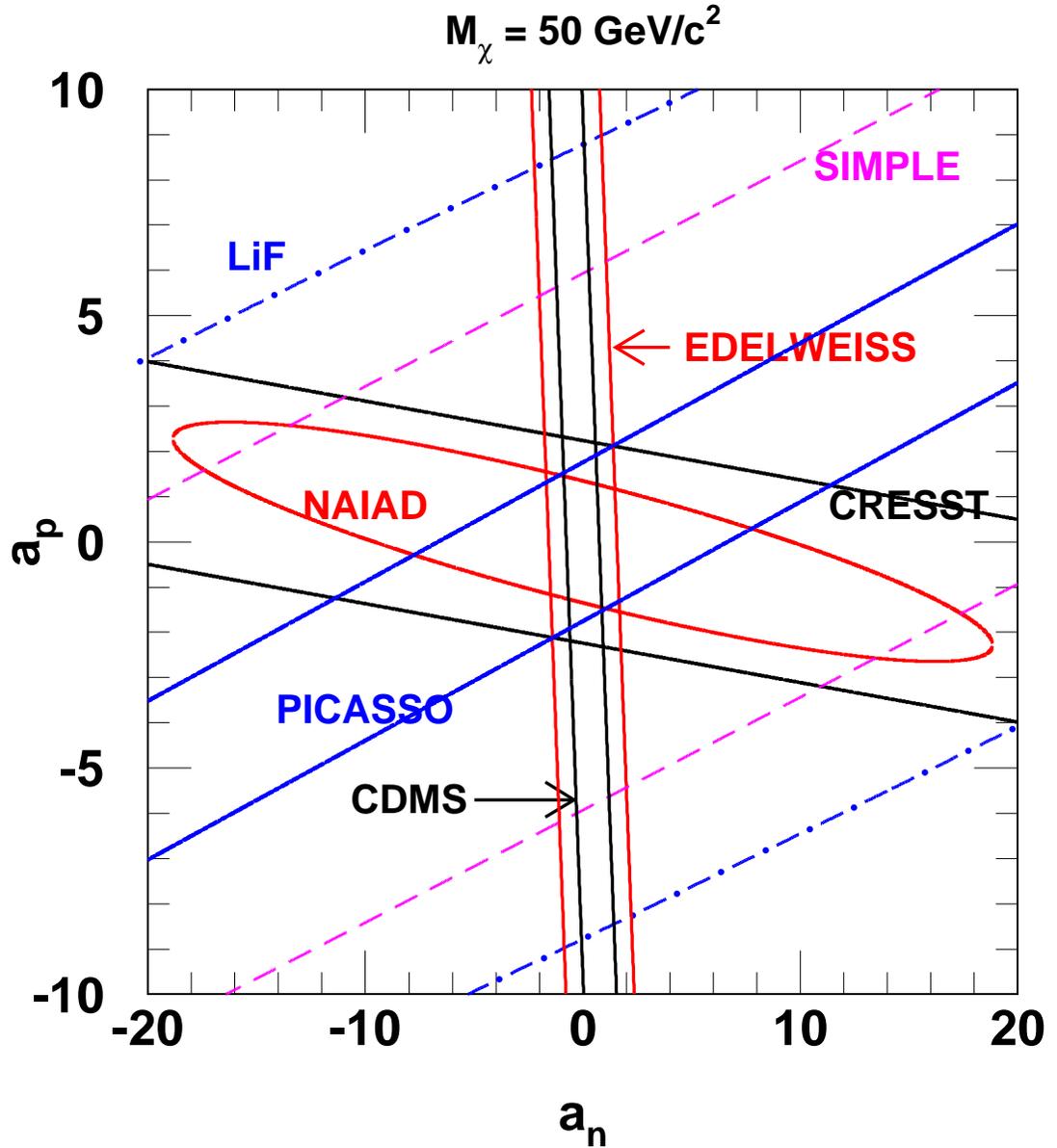
,height=17cm}
\caption[]{
Allowed regions in the $(a_n,a_p)$ plane for
$M_{\chi}=$50 GeV/c$^2$, for the experiments quoted in the text. The
region allowed by each experiment is within the corresponding parallel
straight lines or ellipse.
          } 
\label{fig-3} 
\end{figure}

\newpage

\begin{figure}[tbh]
\epsfig{file=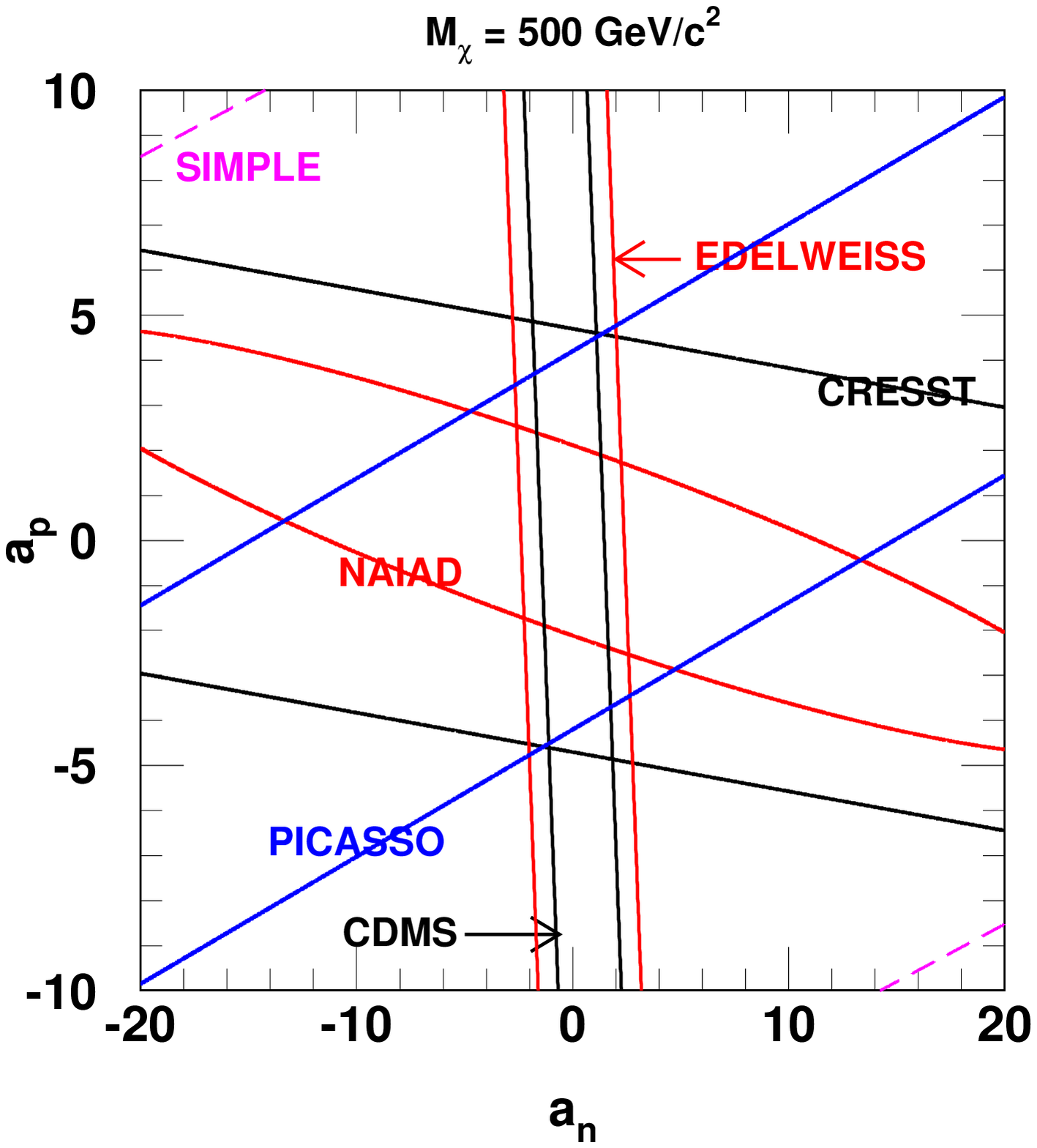
,height=17cm}
\caption[]{
The same as Fig. 3  for $M_{\chi}=$ 500 GeV/c$^2$.
          } 
\label{fig-4} 
\end{figure}


\begin{thebibliography}{99}
%
\bibitem{tovey}
  D.R. Tovey {\it et al.}, Phys. Lett. B {\bf 488} (2000) 17.
\bibitem{jung}
  G. Jungman, M. Kamionkowski and K. Griest, Physics Reports {\bf 267}
  (1996) 195.   
\bibitem{igex}
  A. Morales {\it et al.}, Phys. Lett. B {\bf 532} (2002) 8.  
\bibitem{edw1}
  A. Benoit {\it et al.}, Phys. Lett. B {\bf 545} (2002) 43.
\bibitem{CDMS}
  D.S. Akerib {\it et al.}, Phys. Rev. Lett. {\bf 93} (2004) 211301
\bibitem{cresst}
  G. Angloher {\it et al.}, arXiv:astro-ph/0408006, accepted for publication in Astropart. Phys.
\bibitem{good}
  M.W. Goodman and E. Witten, Phys. Rev. D {\bf 31} (1986) 3059.
\bibitem{giuliani}
  F. Giuliani, Phys. Rev. Lett. {\bf 93} (2004) 161301 
\bibitem{savage}
  C. Savage, P. Gondolo and K. Freese, Phys. Rev. D {\bf 70} (2004) 123513. 
\bibitem{bednyakov2}
  V.A. Bednyakov and H.V. Klapdor-Kleingrothaus, Phys. Rev. D {\bf 70} (2004) 096006.
\bibitem{edw0}
  A. Benoit {\it et al.}, Phys. Lett. B {\bf 513} (2001) 15.
\bibitem{edw2}
  O. Martineau {\it et al.}, Nucl. Instr. Meth. A {\bf 530} (2004) 426
\bibitem{edw4}
  V. Sanglard {\it et al.}, arXiv:astro-ph/0503265, submitted to Phys. Rev. D
\bibitem{yellin} 
  S. Yellin, Phys. Rev. D {\bf 66} (2002) 032005
\bibitem{dimitrov}
  V.I. Dimitrov, J. Engel and S. Pittel, Phys. Rev. D {\bf 51} (1995) R291.
\bibitem{bednyakov}
  V.A. Bednyakov and F. \v Simkovic, arXiv:hep-ph/0406218.
\bibitem{ressel}
  T. Ressel {\it et al.}, Phys. Rev. D {\bf 48} (1993) 5519. 
\bibitem{ahmed}
  B. Ahmed {\it et al.}, Astropart. Phys. {\bf 19} (2003) 691. 
\bibitem{miuchi}
  K. Miuchi {\it et al.}, Astropart. Phys. {\bf 19} (2003) 135.
\bibitem{anglo}
  G. Angloher {\it et al.}, Astropart. Phys. {\bf 18} (2002) 43. 
\bibitem{girard}
  F. Giuliani and T.A. Girard, Phys. Lett. B {\bf 588} (2004) 151.
\bibitem{picasso}
  M. Barnab\'e-Heider {\it et al.}, arXiv:hep-ex/0502028, submitted to Phys. Lett. B
\bibitem{engel}
  J. Engel {\it et al.}, Phys. Rev. C {\bf 52} (1995) 2216. 
\bibitem{bernabei}
  R. Bernabei {\it et al.}, Phys. Lett. B {\bf 436} (1998) 379.
\bibitem{arneodo}
  F. Arneodo {\it et al.}, Nucl. Instr. Meth. A {\bf 449} (2000) 147. 
\bibitem{akimov}
  D. Akimov {\it et al.}, Phys. Lett. B {\bf 524} (2002) 245. 
\bibitem{superK}
  S. Desai {\it et al.}, Phys. Rev. D {\bf 70} (2004) 083523. 
\bibitem{baksan}
  O.V. Suvorova, arXiv:hep-ph/9911415.  
\bibitem{ellis}
  J. Ellis {\it et al.}, Eur. Phys. J. C {\bf 24} (2002) 311.
\bibitem{profumo}
  S. Profumo and C.E. Yaguna, Phys. Rev. D {\bf 70} (2004) 095004. 
\bibitem{edw5}
  P.C.F. Di Stefano {\it et al.}, in: Proc. 6th UCLA Symposium on Sources and 
Detection of Dark Matter and Dark Energy in the Universe, 
February 18-20, 2004, Marina del Rey, California, to be published in New
Astronomy Reviews
\end{thebibliography}
\end{document}